\documentstyle[aaspp]{article}

\def\stacksymbols #1#2#3#4{\def\theguybelow{#2}
	\def\verticalposition{\lower#3pt}
	\def\spacingwithinsymbol{\baselineskip0pt\lineskip#4pt}
	\mathrel{\mathpalette\intermediary#1}}
\def\intermediary #1#2{\verticalposition\vbox{\spacingwithinsymbol
	\everycr={}\tabskip0pt
	\halign{$\mathsurround0pt#1\hfil##\hfil$\crcr#2\crcr
		\theguybelow\crcr}}}
\def\lta{\stacksymbols{<}{\sim}{2.5}{.2}}
\def\gta{\stacksymbols{>}{\sim}{3}{.5}}

\begin{document}
\title{EVOLUTION OF INTERSTELLAR GAS IN RAPIDLY ROTATING 
ELLIPTICAL GALAXIES: FORMATION OF DISKS$^1$}

\author{Fabrizio Brighenti$^{2,3}$ and William G. Mathews$^3$}

\affil{$^2$Dipartimento di Astronomia,
Universit\`a di Bologna,
via Zamboni 33,
Bologna 40126, Italy\\
brighenti@astbo3.bo.astro.it}

\affil{$^3$University of California Observatories/Lick Observatory,
Board of Studies in Astronomy and Astrophysics,
University of California, Santa Cruz, CA 95064\\
mathews@lick.ucsc.edu}

\vskip .2in

\begin{abstract}

We describe the evolution of interstellar gas in a family of low
luminosity elliptical galaxies all having $M_B = -20$ but with
different degrees of flattening (E0, E2, and E6) and two current
supernova rates, SNu = 0.01 and 0.04.  The galaxies are composed of
90 percent dark matter, are rotationally flattened and have isotropic
stellar velocity dispersions.

The soft X-ray luminosity of the hot interstellar gas after evolving
for 15 Gyrs decreases dramatically with increasing galactic rotation.
As the rotating hot interstellar gas loses energy in the galactic
potential, it cools onto a large disk.  The outer radius of the disk
can be much reduced by increasing the supernova rate which drives a
gentle galactic wind transporting high angular momentum gas out of the
galaxy. The total mass of cooled disk gas is less sensitive to the
supernova rate.

Although the hot interstellar gas may be difficult to observe in
rotating low-luminosity ellipticals, the cooled disk gas can be
observed (i) in optical line emission since part of the cooled disk
gas is photoionized by stellar UV and (ii) in the optical continuum,
assuming the colder disk gas forms into luminous stars. The mass of
HII gas ($\sim 10^8$ $M_{\odot}$) may be much greater than previously
realized since rotationally supported, low density HII contributes
little to the global optical line emission. We interpret the stellar
disks that are common (or ubiquitous) in low luminosity ellipticals as
stars that have formed in the cold disk gas. The total mass of cold
disk gas available for star formation is similar to the masses of
stellar disks observed. The high stellar H$\beta$ photometric index
observed in disky ellipticals can be understood by combining the light
of young disk stellar populations with that of the old bulge
population.

\noindent
{\it Subject headings}: galaxies: evolution -- galaxies: disks --
galaxies: interstellar gas

\end{abstract}


\section{INTRODUCTION}

In a previous paper we discussed the evolution and X-ray appearance of
hot interstellar gas in a family of massive, slowly rotating, isolated
elliptical galaxies having different ellipticities and rotation rates
(Brighenti \& Mathews 1996, a.k.a. Paper 1).  When large, non-rotating
ellipticals are flattened by anisotropic stellar velocities we found
that the X-ray images are rather insensitive to variations in the
galactic ellipticity.  However, when a small rotation is introduced
the X-ray images become significantly flatter than the optical image
when viewed perpendicular to the axis of rotation.  As the hot gas
flows inward in the galactic ``cooling flow,'' it forms a large disk
comparable in radius to the effective radius and spins up to the local
disk-plane circular velocity which can be quite large for massive
ellipticals, $\sim 500$ km s$^{-1}$, a velocity that should be
resolved with AXAF.  After evolving from 1 Gyr, when early galactic
winds are assumed to have subsided, to 15 Gyrs, about ten percent of
the total baryonic mass of the galaxy cooled into this large disk.
The final disposition of this cooled gas is uncertain; low-mass star
formation is one possibility but a significant amount of gas may
remain as low density HII gas ionized by the stellar UV.

Now we wish to examine the evolution and X-ray appearance of
interstellar gas in a second type of elliptical galaxy having a lower
luminosity and larger rotation.  In recent years it has become
apparent that the stellar properties of elliptical galaxies divide
into two types: ellipticals of high luminosity having (modestly)
triaxial shapes, boxy isophotes, more pronounced central cores, less
rotation and (limited) flattening by anisotropic stellar dispersion;
by contrast, ellipticals of low luminosity generally have disky
isophotes, dense central stellar cusps, rotationally induced
flattening, approximately axisymmetric geometry and isotropic stellar
velocities (Davies et al. 1983; Nieto et al. 1991; Kormendy \& Bender
1996; Tremblay \& Merritt 1996; Faber et al. 1997) The changeover in
elliptical properties occurs over a range of luminosities, $M_V \sim
-20$ to $-22$ where both types coexist.  Both high and low luminosity
ellipticals coexist on the fundamental plane (Faber et al.  1997).

Our interest in the evolution of hot interstellar gas in low
luminosity ellipticals has been motivated in part by the prospect of
better X-ray observations that will be provided by AXAF in the near
future.  In low luminosity ellipticals the X-ray emission from the hot
gas is likely to be masked by the collective emission from (low-mass)
X-ray binary stars (e.g. Kim, Fabbiano \& Trinchieri 1992).  However,
the stellar X-ray component is expected to have the same surface brightness
distribution as the optical image so it should be possible in
principle to subtract the stellar contribution from high quality AXAF
images, leaving only the emission of the hot gas.  
Although such a subtraction would be difficult because of the low 
x-ray luminosity of the gas, it would be helped by the different
surface brightness distributions expected for stars and gas (see
below), by consistent x-ray surface brightness fluctuations anticipated 
from the stellar
component, and by the dissimilar x-ray spectra of stars and gas. 
The characteristic
stellar velocity dispersion (stellar temperature) in ellipticals
decreases with total stellar mass, $T_* \propto
\sigma_*^2 \propto M_{*t}^{0.32}$ (Faber et al. 1997) so the temperature of
thermalized interstellar gas in virial equilibrium should also
decrease with galactic mass and luminosity.  However, the galactic
binding energy is also proportional to $\sigma_*^2$ so any additional
source of energy in the interstellar gas, such as supernova
explosions, may cause the interstellar gas in low luminosity
ellipticals to become unbound and flow out of the galaxy as a wind.
Little is known at present about the X-ray emission from rapidly
rotating, coreless, disky, low luminosity ellipticals; NGC 4697 may be
the only such elliptical with measured X-ray fluxes (Fabbiano, Kim, \&
Trinchieri 1992). While the evidence for hot gas is currently weak or
nonexistent, more may be learned in the near future.

The gas-dynamical history of low luminosity, disky ellipticals is also
of interest because the stellar disks may result from star formation
in the disk of cold gas that is a natural product of rotating cooling
flows.  The sense of rotation of the stellar disks is identical to
that of the bulge component, supporting an internal origin for the
stellar disks rather than formation in a merging event.  Recently de
Jong \& Davies (1997) have reported that disky ellipticals have higher
H$\beta$ photometric indices, indicating the presence of youthful
stars.  They suggest that stars in the disk component, if sufficiently
young, could account for the apparent youthful age of the entire
galaxy as determined by the population studies of Worthey (1994) and
others.  It is of some interest therefore to explore the possibility
that star formation has occurred in the disks of cold gas that are
expected in rapidly rotating ellipticals.

Finally, the large rotationally supported disks of cooled gas are 
exposed to ultraviolet ionizing radiation present in galactic
starlight.  This raises the interesting possibility that the ionized
component of these disks can be (or has been) observed in faint
optical line emission.  We find that surprisingly large masses ($\sim
10^9$ $M_{\odot}$) of warm ($T \sim 10^4$ K), low density HII gas can
reside in the outer parts of these gaseous disks where is it supported
by rotation at the local galactic circular velocity and in pressure
equilibrium with the ambient hot interstellar gas.  The density of
most of this HII gas is very low and radiates line emission extremely
weakly.  The presence of this warm component, which has not previously
been discussed, eases somewhat the mystery regarding the ultimate
disposition of galactic cooling flow gas after it cools.

In the following we discuss the evolution of hot interstellar gas in a
family of six low luminosity, ellipticals having different
ellipticities, angular momenta and supernova heating rates.  As in
Paper 1 we consider the rotating galaxies to be isolated; gas inflow
or ram pressure effects expected in a cluster environment are not
considered here.  We find that the radial extent of the disks, the
mass of gas entering the disks, and the current mass of hot
interstellar gas throughout the galaxies are all very sensitive to the
assumed Type Ia supernova rate, assuming that the supernova energy is
shared throughout the interstellar gas.  In the most favorable
circumstances for disk formation in rapidly rotating galaxies, the
disks are very extended.

\section{GALACTIC MODELS}

A variety of double power law density models
has been proposed for spherical elliptical galaxies (Jaffe 
1983; Hernquist 1990; Dehnen 1993; Tremaine et al. 1994).  
Some of these provide excellent global approximations to 
de Vaucouleurs profiles while others (Faber et al. 1997) 
are designed to match 
central density profiles observed with HST.
However, 
apart from their success in fitting the projected 
surface brightness distribution,
the lack of dark matter and infinite extent of 
these double power law models make them inappropriate
for determining stellar velocities for our models.
On the other hand we find that the projected 
surface brightness of King 
profiles are adequate approximations to de Vaucouleurs profiles,
certainly for our purposes here, 
and also have much mathematical simplicity particularly when
extended to include oblate galaxies.
 
For these reasons we have decided to retain the models used in 
our previous paper (Brighenti \& Mathews 1996, Paper 1) on 
more luminous ellipticals:
simplified King profiles 
for the stellar density and pseudo-isothermal structures for 
the dark matter distribution, i.e.
$$\rho_*(m_o~^2) = \rho_{o*}(1 + (R_t/R_{c*})^2 m_o^2)^{-3/2}
~~~~~~~
\rho_h(m_o~^2) = \rho_{oh}(1 + (R_t/R_{ch})^2 m_o^2)^{-1}.$$
The two density distributions depend on a single
parameter $m_o^2 = (R/R_t)^2 + (z/z_t)^2$
and are both truncated at some large elliptical 
surface defined by $R_t$ 
and $z_t = R_t (1 - e^2)^{1/2}$ where $e$ is the eccentricity 
of both the stellar and dark matter.
We generate a family of equal-mass elliptical En galaxies all 
having the same $\rho_{o*}$ and $\rho_{oh}$ by scaling the $R-$core radius
according to $R_{c*} = R^{(s)}_{c*}(1 - 0.1n)^{-1/3}$ where
$R^{(s)}_{c*}$
is the core radius of the spherical galaxy.
The detailed procedure used in constructing the models 
and the total galactic 
potential (both within and beyond the galactic mass distribution) 
are described in Paper 1.
As in Paper 1 we solve the two dimensional Jeans equations to 
find the stellar temperature
$$T_*(R,z) = {\mu M \over 3 k_B} (2 \sigma^2 + \sigma_{\phi}^2 )$$
at every position in the galaxy.
Here 
$\sigma^2 = \sigma_R^2 = \sigma_z^2$ 
is the stellar velocity dispersion in the meridional plane 
and $\sigma_{\phi}^2$ is the random stellar dispersion in the 
azimuthal direction.
In Paper 1 we employed the Satoh decomposition of the mean 
square azimuthal speed into a random dispersion and a systematic
rotation:
$$\overline{v_{\phi}~^2} = \sigma_{\phi}~^2 + \overline{v_{\phi}}^2$$
where
$$\sigma_{\phi}^2 \equiv \overline{v_{\phi}~^2} -
\overline{v_{\phi}}^2 = k^2 \sigma^2
+( 1 - k^2) \overline{v_{\phi}~^2}$$ (Satoh 1980).  In general
$k^2(R,z)$ is an unknown function of both spatial coordinates, but is
usually regarded as constant throughout the galaxy. In this current
paper we consider only galaxies that
are rotationally flattened, i.e. $k^2 = 1$ and which therefore must
have have isotropic velocity ellipsoids.  The evaluation of the
potential and stellar velocity dispersion simplifies considerably in
the spherical limit, see Paper 1 for further details.

We use an extended Hubble nomenclature to describe our models:
EnA;k$^2$ where En is the normal Hubble designation with $n = 10[1 - (1 -
e^2 
)^{1/2}]$, $k$ is the Satoh parameter and A = H or L is an
additional optional distinction that characterizes the 
``high'' or ``low'' supernova rate assumed
(as explained below).  We consider here a family of galactic
models all having the same luminosity, central densities, 
and total mass but with
different degrees of rotationally induced flattening.  
In Table 1 we list the parameters of
the spherical member of the family, E0;0.
The basic parameters that
define the stellar distribution, $R_e$, $L_B$ and $\sigma_*$ are
chosen so that the galaxy lies on the fundamental plane (Tsai \&
Mathews 1995; Paper 1).  These parameters are also nicely consistent
with recent fits to HST observations (Faber et al. 1997) with one
exception: our stellar core radius $R_{c*} = 63$ pc is about
ten times larger than upper limits for 
possible cores in ``power-law'' galaxies of the
same absolute magnitude.  If we decrease the core radius to be 
consistent with
the HST upper limits while also maintaining the 
Tsai-Mathews fundamental plane
relationships, we find that the overall luminosity of the galaxy
becomes too faint to be of much interest ($M_V \gta -18$).  
Nevertheless, 
the galaxy described in Table 1 is entirely satisfactory
for our models of the global large-scale
evolution of interstellar gas throughout the galaxy
in which the stellar core 
region has no influence and is poorly resolved by the 
hydrodynamic grid.  Finally we note that the central density and core
radius of the dark matter component have been chosen so that the total
dark mass is 9 times that in the stellar component and its
distribution is such that dark matter has little influence on stellar
velocities within $\sim 1$ effective radius.

Figure 1 shows the distribution of stellar temperature $T_*(R,z)$ and
azimuthal velocity $v_{*\phi}(R,z)$ in a quadrant of the E2;1 member
of the galactic family corresponding to the E0;0 galaxy in Table 1.
At the outer edge of the galaxy the stellar velocity dispersions in
the meridional plane must vanish since no stars can move beyond this
surface; however, when $k=1$ the dispersion is isotropic and the
azimuthal dispersion $\sigma_{\phi}^2$ (and $T_*$) must also vanish at
the outer galactic boundary.  Consequently, the contours of $T_*(R,z)$
are much more elliptical than those of the (slowly rotating) E2;0.25
galaxy shown in Paper 1.

The azimuthal stellar velocity $v_{*\phi}(R,0)$ and circular velocity
on the equatorial plane are illustrated in Figure 2.  The rather
strong maximum in $v_{*\phi}(R,0)$ at a few hundred parsecs from the
galactic center, particularly for the E6;1 galaxy, may differ from
stellar rotation curves observed in some low luminosity ellipticals
[e.g. NGC 4697 observed by Binney, Davies, and Illingworth (1990)],
but stellar rotation curves are unavailable for most of the low
luminosity ellipticals known to have power-law density profiles.  In
any case, the global interstellar gas dynamics are not strongly
influenced by this maximum in $v_{*\phi}(R,0)$.  As the hot
interstellar gas cools and flows inward, it is expected to spin up to
the local circular velocity $v_{circ}(R)$ and become rotationally
supported.

\clearpage
\section{GAS DYNAMICS}

The gas dynamical equations are identical to those used in Paper 1.
Interstellar gas is created by normal mass loss from an evolving
population of old stars at a rate $$\alpha_*(t) \rho_*= \alpha(t_n)
(t/t_n)^{-1.3} \rho_* ~~~ {\rm gm}~{\rm cm}^{-3}~{\rm s}^{-1}$$ where
$t_n = 15$ Gyr represents the present time and $\alpha(t_n) = 5.4
\times 10^{-20}$ s$^{-1}$ (Mathews 1989).  The contribution of gas
from Type Ia supernovae to the interstellar mass is negligible
($\alpha_{sn} \ll \alpha_{*}$) but is more energetic and iron-rich.
The energy generated at early times by Type II supernovae is expected
to expel most of the interstellar gas created prior to some time which
we adopt as 1 Gyr.  Between 1 and 15 Gyrs about one tenth of the
stellar mass remaining at $t = 1$ Gyr enters the interstellar medium.
For simplicity, we do not vary the galactic potential during this time
interval to account for this small stellar mass loss.

The temperature of the interstellar gas is similar to the local
stellar temperature since both components are in hydrostatic
equilibrium in the galactic potential, although the gas can also be
heated by supernova explosions.  The characteristic source-term
temperature for the gas is $$T_o = (\alpha_* T_* + \alpha_{sn}
T_{sn})/\alpha_*$$ where $\alpha_{sn} \ll \alpha_{*}$ is assumed.
Heating by supernova explosions is given by $$\alpha_{sn} T_{sn} =
2.13 \times 10^{-8}~ {\rm SNu}(t)~ (E_{sn}/10^{51} {\rm ergs})~
h^{-1.7}~ (L_B/L_{B \odot})^{-0.35} ~~~ {\rm K}~{\rm s}^{-1}$$ 
(see Mathews 1996)
where $h \equiv H/100 = 0.75$ is the reduced Hubble constant and we
adopt $E_{sn} = 10^{51}$ ergs as the typical hydrodynamic energy
released in a supernova event.  
The temperature $T_{sn} = m_p E_{sn} / 3 k m_{sn}$ ($m_p =$ proton 
mass) depends on 
the average mass lost per supernova $m_{sn}$. 
The total rate that supernovae supply gas to a galaxy of mass
$M_{*t}$ is $\alpha_{sn} M_{*t} = \nu_{sn} m_{sn}$ where 
$\nu_{sn}$ is the supernova rate in sec$^{-1}$; 
to convert $\nu_{sn}$ to SNu units (number of supernovae per 100 
years from stars of luminosity $L_B = 10^{10} L_{B \odot}$)
we use the mass to light ratio determined by van der Marel (1991),
$M_{*t} / L_B = 2.98 \times 10^{-3} L_B^{0.35} h^{1.7}$ 
where $L_B$ is in $L_{B \odot}$.
The supernova rate is assumed to
decrease with time as a power law: $${\rm SNu}(t) = {\rm SNu}(t_n) (t
/ t_n)^{-1},$$ although there is at present no compelling
observational reason to adopt this form.  We consider two values for the
current supernova rate, ${\rm SNu}(t_n)= 0.01$ and ${\rm
SNu}(t_n)= 0.04$.  These rates are comparable with the most recent
observed value ${\rm SNu}_{obs}(t_n)= 0.12
\pm .06 ~ (h/0.75)^2$ (Turatto, Cappellaro \& Benetti 1994).
We do not expect exact agreement between the observed value ${\rm
SNu}_{obs}(t_n)$ and the factor ${\rm SNu}(t_n)$ in the expression
above for $\alpha_{sn}T_{sn}$ since SNu$(t)$ is only one of several
uncertain parameters that influence $\alpha_{sn}T_{sn}$ in our gas
dynamical models.  With the higher value of ${\rm SNu}$ we
find that the iron abundance in the interstellar gas exceeds those 
observed by the ASCA satellite; possible origins of this discrepancy
have been discussed by Arimoto et al. (1997). For simplicity we
distribute the supernova energy (and iron!) smoothly throughout the
interstellar gas; the accuracy of this often-used assumption is
difficult to access since the initial deposition of supernova energy
is localized in hot interstellar bubbles (Mathews 1990). Since
$\alpha_*(t)$ decreases with time more rapidly than SNu$(t)$, the
influence of supernova heating (and the likelihood of galactic winds)
increases with time in our models.

As in Paper 1 we assume that the galaxy is completely isolated,
allowing hot gas to flow freely beyond the stellar edge.  The gas flow
is regarded as invicid, i.e. the magnetic field is sufficiently large
to drastically reduce the plasma mean free path yet the field is
sufficiently disordered on small scales ($\ll R_e$) not to communicate
stresses over large distances in the interstellar gas. For numerical
solutions describing the evolution of the interstellar gas we used the
NCSA Eulerian hydrocode ZEUS2D, appropriately modified as described in
Paper 1.

\section{COOLING FLOW EVOLUTION}

\subsection{The Spherical E0;0 Galaxy}

The influence of galactic rotation on the nature and evolution of
interstellar gas in ellipticals can be determined by comparison with
otherwise identical spherical, non-rotating models. Properties of the
interstellar gas in the non-rotating E0;0 galaxy at time $t_n = 15$
Gyr are shown in Figure 3.  The interstellar gas flows in a
axisymmetric cylindrical computational grid of 120 $\times$ 120 zones
with gradually increasing zone size; the smallest central zone is half
the stellar core radius. Solutions are initiated at time $t = 1$ Gyr,
when it is assumed that SNII explosions and early galactic winds have
subsided, and continued to 15 Gyrs with two values of the current
supernova rate, SNu$(t_n) = 0.01$ (E0L;0) and SNu$(t_n) = 0.04$
(E0H;0).  The linear excursions of the gas density and temperature at
small radii in Figures 3a and 3b are artifacts of the limited spatial
resolution in the innermost grid points.  The temperature of the gas
is slightly larger than the virial temperature for the solution with
the higher supernova rate (E0H;0) so some gas flows out of the galaxy.
As a result the gas density and X-ray surface brightness in the E0H;0
galaxy are lower than for the E0L;0 galaxy while the gas temperature
in the E0H;0 galaxy is higher, particularly in the outer parts of the
galaxy. The X-ray surface brightness is calculated for the 0.1 - 2.4
keV energy band to approximate ROSAT observations.

The effect of changing the supernova rate is also apparent in Table 2
which lists several global properties of these models.  The mass of
hot gas within the galactic boundary $M_g(hot)$ and (especially) the
total soft X-ray luminosity $L_x$ (0.1 - 2.4 keV) are both reduced
when the supernova rate is increased.  The total amount of cold gas
that accumulates in the galactic core $M_g(cold)$ is also
significantly lower as a result of gas outflow.

The gas velocity in these spherical galaxies is very subsonic
everywhere. As gas loses energy by radiation it sinks subsonically in
the galactic potential and is maintained at approximately constant
temperature by $Pdv$ compression in the galactic potential.
Superimposed on this global flow are low-amplitude, highly subsonic
random motions ($\sim 10 - 20$ km s$^{-1}$). We believe these random
velocities are numerical in origin, but similar small velocities could
be generated in real galaxies by small physical irregularities.  These
small velocities are nevertheless sufficient to prevent the appearance
of a ``galactic drip'' or global thermal instability (Mathews 1997)
that appears when these same calculations are performed with a 1D
spherically symmetric grid.  The implications of these results for the
reality of galactic drips are unclear at present.

\subsection{Cooling Flow Evolution in E2;1 and E6;1 Galaxies}

\subsubsection{Formation of Cold Disks}

As expected from angular momentum conservation, in rotating galaxies
the gas cools into a large, rotationally supported disks .  Gas that
cools below $10^4$ K is held at this temperature throughout the
remainder of the calculation.  The mass of this cold gas in the disk
generally increases with time.  The total vertical column density in
cooled gas is shown at three times in Figures 4a, b, and c for the
disks of the E2L;1, E2H;1, and E6L;1 galaxies respectively.  The E2H;1
elliptical with the higher supernova rate (Fig. 4b) has a much less
extended disk owing to the outflow of (high angular momentum) gas from
the galaxy in a slow wind. At the last time plotted in Figure 4b the
E2H;1 disk has actually become smaller probably as a result of local
heating of gas by supernovae. However, the large disks that form in
the E2L;1 and E6L;1 galaxies are very similar.  In both disks the
outer edge increases with time, similar to the growth pattern of disks
in more slowly rotating ellipticals (Paper 1) and the surface density
increases slowly within the body of the disk, indicating that gas
enters the disk at all radii, not just at the outer edge.  The
similarity of Figures 4a and 4c indicates that the column density
beyond several stellar core radii is almost independent of the degree
of galactic rotational flattening.  The E2L;1 and E6L;1 galaxies both
have power-law, not exponential, disks with surface densities given by
$\Sigma_d \propto R^{-1.93}$.

Even in these maximally rotating galaxies with $k^2 = 1$, gas ejected
from stars moves a considerable distance from its point of origin
toward the rotation axis before settling into the cold disk. 
As gas moves along a streamline toward the disk, it mixes
with locally produced gas having different angular momentum.
In spite of this complication, it is possible to estimate 
the radius $R_d(R,z)$ in the disk where 
mass lost from stars at $R,z$ will eventually cool 
(see Paper 1).
In Figure 5 we plot contours of 
equal $R_d(R,z)$ for the E2:1 galaxy.
These contours 
are (implicit) solutions of $R_d v_{circ}(R_d) = R v_{* \phi} 
(R,z)$, the condition that an element of gas ejected by stars at 
$R,z$ strictly 
preserves its specific angular momentum until it reaches the disk 
at radius $R_d$.
For example, Figure 5 indicates that gas ejected
from stars at $R \sim 20$, $z \sim 20$ kpc has the same 
angular momentum as gas in the disk at $R_d
\approx 4$ kpc.

\subsubsection{Nature of Interstellar Medium at $t_n = 15$ Gyr}

The flow in galaxy E2L;1 shown in Figure 6 exhibits subsonic kinetic
activity and corresponding irregularities in the density contours,
particularly in the outer galaxy.  Within about 10 kpc, however, the
flow toward the central disk is regular and the density contours have
an ellipticity similar to that of the local stars.  However, notice
that a relatively dense and cold region has appeared in the flow at
time $t_n = 15$ Gyr near $R = 40$, $z = 3$ kpc. This cold, transient
lump of gas is falling toward the disk plane where it will eventually
reside.  Such intermittent contributions to the far outer edge of the
cold disk are also evident from Figure 4 and may be a generic feature
in rapidly rotating cooling flows. Similar inhomogeneities have also
appeared in the calculations of D'Ercole \& Ciotti (1997) who are
studying the evolution of hot gas in S0 galaxies.  At time $t = 15$
Gyrs the most rapid rotation above the disk plane, $v_{\phi} \sim 220$
km s$^{-1}$, is near $R \approx 28$ kpc; the full width of emission
line profiles in the X-ray spectrum of this galaxy is $\sim 440$
km s$^{-1}$, within the resolution capability of AXAF.

The irregular velocity field in Figure 6, unlike the laminar flow in
the slowly rotating galaxies described in Paper 1, may result from
shear instabilities that first develop in the low density outer
galaxy. The azimuthal velocity of gas just within the galaxy is driven
somewhat by ejecta from azimuthally streaming stars.  Such an
interaction may enhance shear instabilities at the galactic boundary,
but we have not investigated in detail the physical origin of these
very subsonic velocities.

When the supernova rate is increased by a factor of 4, a subsonic wind
develops throughout most of the galaxy at time $t_n = 15$ Gyr as
shown in Figure 7.  The gas density is lower and its spatial variation
is more regular in this gentle galactic wind than in the flow in
Figure 6.  Since we assume that mass enters the flow at a rate
$\alpha_* \propto t^{-1.3}$ that decreases faster than the supernova
rate, SNu$(t) \propto t^{-1}$, the thermal energy delivered to each
gram of interstellar gas slowly increases with time and the overall
solution tends toward a galactic wind as the calculation proceeds.
Finally, in Figure 8 we show the density and velocity fields at $t_n =
15$ Gyr in the rapidly rotating E6L;1 galaxy.  In such flat galaxies
incipient winds first appear near the outer edge of the equatorial
plane near $R
\sim 80$ kpc.  Winds occur first in the equatorial region because (i) the
(more spherical) gravitational potential is less here and (ii) the
presence of rotation reduces the effective potential.  Notice that
there is another ``lump'' of cooling gas descending toward the
equatorial plane at $R \approx 45$ kpc similar to the cool lump
discussed in Figure 6.

\subsubsection{Soft X-ray Images}

The appearance of the hot interstellar gas when viewed in soft X-rays
(0.1 - 2.4 keV) is shown for three galaxies in Figure 9.  The galactic
breeze driven by supernova heating in the E2H;1 galaxy has a dramatic
effect in circularizing the X-ray image compared to that of the E2L;1
galaxy of the same age.  However, it is unclear if the emission from
the hot gas illustrated in Figure 9 can be observed because it is
masked by the collective emission from low mass X-ray binaries in the
stellar component.  Ideally, with high signal to noise AXAF
observations it would be possible to subtract the soft X-ray
contribution from stars since the stellar surface brightness
distribution is known from optical observations.  Although the
possibility of such a decomposition to find the X-ray emission from
the gas alone was one of the initial motivations for our series of
calculations, the X-ray luminosities of the interstellar gas in the
rotating galaxies in Table 2 may be too small to be detectable. 
The stellar $L_x$ from ellipticals is poorly
known, but Ciotti et al. (1991) suggest $L_{x,*} \approx 1.5
\times 10^{40} (L_B / 10^{11} )$ ergs s$^{-1}$ which for our galaxy
is $L_{x,*} \approx 2 \times 10^{39}$ so the $L_x$ from the
interstellar gas will be a disky component only a few percent of the
total -- a challenging observation even for AXAF.  The X-ray emission
from non-rotating galaxies in Table 2 could be observed quite easily,
but it is unlikely that many low luminosity galaxies are either
spherical or non-rotating (Tremblay \& Merritt 1996).

\section{PHYSICAL NATURE OF THE COOLED GAS}

The final physical configuration of cooled gas in galactic and cluster
cooling flows remains one of the most perplexing unsolved problems in
galactic evolution.  The usual solution is to assume that stars of
very low mass and luminosity form in these high-pressure environments;
there is currently no strong observational evidence against this
hypothesis.  In the following discussion, however, we examine two new
aspects of this problem that arise when galactic rotation is
considered.  First, regardless of the final endstate of the cooled
gas, there will always be some disk gas that cannot form into stars
because it is photoionized by the galactic and intergalactic ionizing
radiation fields.  We discuss below the mass and possible
observability of this ionized gas.  Second, we consider the hypothesis
that relatively young luminous stars have formed in the cooled gaseous
disks and that these disk stars are responsible for the systematically
larger H$\beta$ photometric indices observed in disky ellipticals of
low luminosity.

\subsection{Emission from HII gas at $T \approx 10^4$ K}

Some of the gas that cools onto the rotating disk is kept ionized by
ambient ionizing radiation and remains at $T \approx 10^4$ K, the same
temperature we have adopted for all the cooled gas.  In the inner
parts of our disks the maximum column depth of HII gas that can be
ionized by the available radiation, $N_i$, is much less than the 
total column of cooled disk gas
$N_{tot}$.  Such a layer of ionized gas should be stable toward
forming gravitational irregularities.  However, regions deeper in the
disk than $N_i$ are in reality expected to be colder than $T = 10^4$ K
and are probably unstable to cloud (and star?) formation. If dense
condensations should form in this colder gas, the HII gas
should still be present as a distributed component having a column
density $\sim N_i$.  In the inner disk therefore we expect a column
$N_i$ of HII gas to always be present even if most of the colder disk
gas is highly clumped or stellar.  In addition, our solutions indicate
that a considerable mass of HII gas may be present in the outer disk
provided the supernova rate is low.
Figure 10
shows a plot of the variation of the total column of cold gas
$N_{tot}$ and the maximum amount that can be ionized $N_i$.
Beyond about $R \sim 9$ kpc in the
E2L;1 galaxy all the disk is fully ionized.
We describe below how we have estimated the column $N_i$ and the total
mass of HII gas.  We also discuss the likelihood of detectable optical
line emission from the HII disk.  The ionizing radiation has both
galactic and extra-galactic components.  

Stellar UV flux has been measured in seven ellipticals from 1800
\AA$~$ to the Lyman edge (Ferguson \& Davidsen 1993; Brown, Ferguson, \&
Davidsen 1995) and Binette et al. (1994) have estimated the entire UV
spectral energy distribution and the total galactic ionizing photon
luminosity ${\cal L}_t$ from the collective radiation of post-AGB
stars in an old stellar population. Such estimates are very uncertain
since the evolutionary tracks of highly evolved stars and their
atmospheres are poorly understood. At time $t = 13$ Gyr following 
a single burst of star formation, Binette et al.
find ${\cal L}_t \approx 7.3 \times 10^{40}$ s$^{-1}$
${M_{\odot}}^{-1}$; for our galaxy this is ${\cal L}_t \approx 5.5
\times 10^{51}$ s$^{-1}$.  We have attempted to reconcile this
estimate with the UV observations by averaging values of $\nu
L_{\nu}/(L_B/L_{B,\odot})$ at 1100 \AA$~$ and 1500 \AA$~$ for the six
galaxies observed by Brown, Ferguson, \& Davidsen (1995). Our values
of $L_B$ are based on $h = 0.75$ and the galactic properties listed by
Faber et al. (1997).  Individual values of $\nu
L_{\nu}/(L_B/L_{B,\odot})$ at both wavelengths range over a factor of
about four, but the ratio of the mean values at these two wavelengths
agrees almost exactly with the slope of the UV spectral energy
distribution in this part of the spectrum given in arbitrary units by
Binette et al. (1994).  Using the UV observations at these two wavelengths
to calibrate the 
ionizing continuum of Binette et al. (1994) we find ${\cal L}_t
\approx 2.0 \times 10^{41} (L_B / L_{B,\odot})$ s$^{-1}$ which for our
galaxy (e.g. Table 1) is ${\cal L}_t \approx 2.6 \times 10^{51}$
s$^{-1}$ in good agreement with the estimate of of Binnette et al.  We
shall therefore adopt here a value ${\cal L}_t = 5 \times 10^{51}$
s$^{-1}$ at $t = 15$ Gyr which is slightly less than the value from
Binnette et al. at $t = 13$ Gyr since ${\cal L}_t$ slowly decreases
with time.  Some of this Lyman continuum radiation is consumed
ionizing dense circumstellar gas in newly formed planetary nebulae,
but we expect this fraction to be small since the nebular gas is
swept back along the stellar orbit by ram pressure on time scales
short compared to the total UV lifetime of the central stars (Mathews
1990).  If the total stellar ionizing photon emissivity follows the
starlight, $j_{iph} = j_o [1 + (R/R_{c*})^2]^{-3/2}$, we determine the
coefficient $j_o = 2.4 \times 10^{-12}$ cm$^{-3}$ s$^{-1}$ and the
ionizing photon density at any radius in the galaxy by using the
expressions for the radiation density from Tsai and Mathews (1995).
For simplicity we use the spherical relations of Tsai and Mathews to
estimate the stellar ionizing photon density and ionized column in the
disk of the moderately flattened E2L;1 galaxy.  For the intergalactic
contribution to the ionizing photon density we use the value
$n_{iph,ig} \approx 4
\pm 2.5 \times 10^{-6}$ cm$^{-3}$ (Dove \& Shull 1994) which is
constant throughout the galaxy.

The ionized column along the disk plane $N_{HII} = {\rm min} (N_i,N_{tot})$
is shown in Figure 10 for the E2L;1 galaxy at time $t = 15$ Gyrs.
Beyond the galactic core radius in the E2L;1 galaxy the ionizing
photon density varies as $n_{iph} \approx 3 \times 10^{-3}
R_{kpc}^{-1.5}$ cm$^{-3}$. This stellar ionizing UV dominates the
intergalactic component throughout most of the galactic volume within
$\sim 35$ kpc which includes all of the disk region.  We assume that
the HII gas in the disk is in pressure equilibrium with the local hot
interstellar gas and has temperature $T \approx 10^4$ K everywhere.
Since the temperature of the cooled gas in our computations is
constrained to $T = 10^4$, it already has the appropriate density for
the HII gas although the density in our cold isothermal disks
decreases somewhat away from the disk midplane.  Beyond the galactic
core radius $R_{c*}$ along the midplane of the disk the density of HII
gas decreases sharply with radius: $n_{i} \approx 50 R_{kpc}^{-3.0}$
cm$^{-3}$.  Estimates of the total mass of HII gas in ellipticals
based on a single assumed density are therefore misleading and
incorrect.  In the dense inner parts of the disk only a small fraction
of the cold disk gas can be ionized but beyond about 9 kpc all the
disk gas is ionized.  The total mass of ionized gas is very large, $8
\times 10^{8}$ $M_{\odot}$, more than ten percent of the total amount
of gas that cooled in the disk since 1 Gyr (Table 2).  The mass of HII
gas within radius R along the disk $M_{i}(<R) \approx 8.7 \times 10^5
R_{kpc}^{2.62}$ $M_{\odot}$ (valid from $R_{kpc} = 1$ to 12.6)
increases dramatically toward the outer parts of the disk where most
of the HII mass resides.  The total column density of all the gas that
has cooled after $t = 15$ Gyr is large, $N_{tot} \approx 5.8 \times
10^{-21} ~ R_{kpc}^{-2.25}$ cm$^{-2}$ (valid from $R_{kpc} = 0.1$ to
10).  This gas cannot be smoothly distributed in the disk (i.e. not in
form of clouds or stars) since it would be easily observed by X-ray
absorption out to 2.5 kpc where $N_{tot} \sim 10^{21}$ cm$^{-2}$.
However, when self-gravity of the HII gas is included we 
find that it is gravitationally 
stable against collapse at every radius in the disk so 
the HII gas will not form into clouds.

The total H$\beta$ luminosity from the disk in our E2L;1 galaxy, $L_{H
\beta} \approx 8 \times 10^{39}$ ergs s$^{-1}$, corresponds to a flux $F_{H
\beta,20} \approx 1.6 \times 10^{-13}$ ergs cm$^{-2}$ s$^{-1}$ when
viewed at a distance of 20 Mpc.  This luminosity and flux are
comparable to those observed in ellipticals of similar optical
brightness (e.g. Trinchieri
\& di Serego Alighieri 1991) so we can expect that ionized disks
contribute a substantial fraction of the total line emission from
early type galaxies.  Because of the strong variation of HII gas
density and ionizing photon density with galactic radius, only about 4
percent of the total mass of HII in the dense central regions, $R \lta
3.2$ kpc, contributes half of the total H$\beta$ luminosity while half
of the total HII mass lies far out in the galaxy beyond $\sim 9.3$ kpc
where all the disk gas is ionized.  This low density HII gas at
$\gta 9$ kpc emits only $\sim 10$ percent of the total $L_{H \beta}$
over a vast surface area and may be undetectable with present technology.
However, the total Balmer line luminosity $L_{H \beta} \approx 8
\times 10^{39}$ ergs s$^{-1}$ far exceeds the minimum luminosity expected in
galactic cooling flows due to the recombination of gas cooling for the
first time; for the E2L;1 galaxy this latter value is $L_{H\beta,min}
\approx 3 \times 10^{36}$ ergs s$^{-1}$ at time $t_n = 15$ Gyr.
Finally we note that $L_{H \beta}$ is about 45 times larger than the
soft X-ray luminosity $L_x$ from the E2L;1 galaxy listed in Table 2 and
therefore represents a significantly larger fraction of the overall
galactic radiation budget.

We have examined in a similar manner the HII mass and line luminosity
in the other galaxies listed in Table 2.  The radial size of the disk
in the E2H;1 galaxy, $R \approx 0.5$ kpc, is greatly reduced by the
higher supernova heating rate assumed.  The total mass of HII gas
expected from the available ionizing photons is much smaller, $M_i
\approx 1.6
\times 10^4$ $M_{\odot}$, but $L_{H\beta} \approx 1.3 \times 10^{39}$
erg s$^{-1}$ is only six times less than that of the E2L;1 galaxy
since most of the H$\beta$ emission is from dense gas in the inner
disk.  The corresponding flux at 20 Mpc is $F_{H\beta,20} \approx 2.7
\times 10^{-14}$ ergs cm$^{-2}$ s$^{-1}$.  The relative insensitivity 
of the 
H$\beta$ luminosity to the physical size of the disk is interesting 
because it cannot explain the enormous range in 
$L_{H\beta}/L_B$ observed; for example Trinchieri
\& di Serego Alighieri (1991) find that $L_{H\alpha}/L_B$ varies over 
a factor of $300$ among ellipticals having $L_B \sim 3 \times 10^{10}$
$L_{B \odot}$. The explanation for this large spread in Balmer
luminosities
could lie in varying magnetic field strengths in the galactic centers;
a field of $H \gta 10^{-3}$ gauss in the dense HII gas near the 
center of our E2L;1 galaxy could sharply reduce the gas density and the 
H$\beta$ emissivity.
The E6L;1
galaxy has an extended disk with the following properties: $M_i
\approx 1.8 \times 10^9$ $M_{\odot}$, $L_{H\beta} \approx 9 \times
10^{39}$ erg s$^{-1}$, and $F_{H\beta,20} \approx 2 \times 10^{-13}$
ergs cm$^{-2}$ s$^{-1}$.

What evidence exists for the presence of HII disks in ellipticals?
Smooth, extended images of ellipticals in optical emission lines (e.
g. Trinchieri \& di Serego Alighieri 1991; Macchetto et al 1996) do
not necessarily require emission from ionized disks and indeed some
HII emission from young planetary nebulae and other ionized stellar
ejecta is expected throughout the galactic volume (Mathews 1990).
Unfortunately it is difficult to estimate the precise amount of HII
gas that has been recently ejected from stars but has not yet entered
the hot phase since the rate of this transition depends on the
influence of the magnetic field configuration on thermal conductivity.
Buson et al.  (1994) present optical line images of a sample of
ellipticals known to be unusually bright in optical emission lines.
These ellipticals also have LINER spectra perhaps suggesting an
additional non-stellar ionization source although the line emission
symmetry does not have the biconical pattern often observed in Seyfert
galaxies. Buson et al.  typically find smooth H$\alpha$ + [NII] images
within $\sim 0.3 R_e$ that are extended along the major axis (NGC
1395, NGC 1453, NGC 2974, IC 1459).  The kinematics of the HII in
these ellipticals indicates regular, organized rotation around the
galactic center (Zeilinger et al. 1996); this rotation is consistent
with disk emission but some (smaller) rotation would also be present
in line emission from recent stellar ejecta distributed throughout the
galactic cooling flow.  In general the emission line surface density
in the HII ``disks'' observed by Buson et al. is confined to the
central regions and drops faster with radius than the stellar light.
In a few ellipticals Buson et al.  find that the major axis of the HII
gas is misaligned with the optical axis of the stellar image (NGC
3962) or is counter-rotating (NGC 6868, NGC 7097); in these galaxies
the HII disks are evidently disturbed or created possibly as a result
of a recent merger event.

Several straightforward observations would help clarify the
contributions of disks to the HII emission from ellipticals. The disk
component is expected to have a higher ellipticity than the stellar
image and this difference should be most pronounced in ellipticals
having high diskyness or $V/\sigma$ that are as a class viewed more
nearly perpendicular to the axis of rotation.  In low luminosity
ellipticals of high $V/\sigma$ the HII rotation velocity should be
representative of the circular velocity on the disk plane which is
always greater than the rotation of the bulk of the galactic stars.


\subsection{H$\beta$ Index: Evidence of Young Disk Stars}

The H$\beta$ index is one of the many observable characteristics that
distinguish elliptical galaxies of high and low luminosity.  This
photometric index measures the strength of Balmer line absorption in
the galactic spectrum and is the principal observational means of
breaking the ``age-metallicity degeneracy'' to reveal the presence of
a subpopulation of youthful stars (Gonzalez 1993; Worthey 1994; Faber
et al 1995; Worthey et al. 1995). Recently de Jong \& Davies (1997)
have shown that H$\beta$ is systematically larger in (low-luminosity)
disky ellipticals (Kormendy \& Bender 1996); they also conjectured
that the young stellar subpopulation may be the stars in the disk.
Using Guy Worthey's stellar population models, de Jong \& Davies
combined the H$\beta$ indices from a young (2 Gyr) stellar disk
containing about 10 percent of the total galactic stellar mass with a
much older component (12 Gyr) representing the bulk of the old stellar
population in the galaxy. They found that the combined H$\beta$ index
of the disk and bulge is similar to the higher H$\beta$
indices observed within $R_e/2$ in disky ellipticals.

We have described here how cold gas gradually collects in a
rotationally supported disk during the slow evolution of the galactic
interstellar medium. This disk provides an ideal environment to form
a younger population of disk stars. Often in theoretical studies
of cooling flows it is assumed that only stars of very low mass form
in the cooled gas expelled from the older generation of galactic
stars.  We shall assume instead that luminous, moderately massive
stars form in the cold disk gas but the stellar masses in the young disk
population must not extend above about 8 $M_{\odot}$ since no Type II
supernova have been observed in ellipticals.  To test this possibility
more quantitatively, we have computed stellar H$\beta$ indices
assuming that the disk consists of a number of discrete population
bursts having mass $\Delta M_i(t_i)$ and age $t_{i}$. These quantized
bursts are listed in Table 3 for the E2L;1 and E2H;1 galactic models.
The burst masses and that of the main galactic bulge are all evaluated
within $R_e/2$ for comparison with the observed values of de Jong \&
Davies (1997).  For simplicity we assume all stellar populations have
solar abundance.

After creating a variety of stellar population models using Guy
Worthey's website (http://www.astro.lsa.umich,edu/users/worthey/), it
became clear that the H$\beta$ index is rather insensitive to the
slope or precise mass cutoffs of the IMF but does depend strongly on
the ages of the youngest stars considered.  Stars younger than the
youngest stellar population available at the Worthey website (1 Gyr)
can contribute significantly to the overall H$\beta$.  In view of
this, Guy Worthey has generously provided us with a version of his
program based on Padova evolution tracks that allows stellar
populations as young as 0.4 Gyr. For the same stellar population, 
H$\beta$ indices based on Padova evolution tracks are $\sim 0.1$
lower than the evolutionary tracks used at the website.  The IMF 
we use here has a Salpeter slope and 
is terminated at mass limits $m_{high} = 10$ and $m_{low} = 0.1$
$M_{\odot}$. For the population ages we consider (Table 3), stars near
the high mass limit do not contribute to the H$\beta$ index, they are
only relevant to estimate the remnant mass. Our remnant masses are
based on evolution tracks currently available at the Worthey website
program.

In Table 4 we list the H$\beta$ index of the main galactic bulge, the
disk component and the two populations combined all evaluated within
$R_e/2$. The H$\beta$ indices are slightly overestimated since we have
only considered the stars within a physical radius of $R_e/2$, not
what would be observed in projection at this radius through the entire
galaxy. On the other hand the indices are underestimated since we (i)
have not considered the contribution from stars younger than 0.4 Gyr
and (ii) assume that there is no time delay between the appearance of
cold gas in the disk and star formation. In any case, the total
H$\beta$ indices listed in Table 4 for galaxy plus disk in the E2L;1
and E2H;1 models nicely span the range of observed values given by de
Jong \& Davies, H$\beta \approx 1.7 - 2.1$.  This greatly supports the
hypothesis that the disk stars are simply a natural outcome of cooling
flow evolution.  There is also some observational evidence that the
disks have higher H$\beta$ indices than the rest of the elliptical
(Scorza \& Bender 1996), but this needs further confirmation.

Our disks are expected to have power-law, not exponential space
density and surface brightness profiles.  The surface brightness
distribution observed in stellar disks of very flattened 
disky ellipticals
(E4 - E6) deviates from exponential toward power law distributions
(Scorza \& Bender 1995). In disky ellipticals the angular
momenta of bulge and disk are always parallel (Scorza \& Bender 1995;
1996); this is consistent with star formation in cooling flow disks
and rules out a disk origin from random mergers.  Finally, we note
that our disks are formed in ellipticals that are completely supported
by rotation with fully isotropic stellar velocities ($k^2 = 1)$. If
some anisotropy were included ($k^2 < 1$), the resulting disks would
be less extended with more steeply varying surface brightness in the
resulting stars.

\section{DISCUSSION AND CONCLUSIONS}

Hot interstellar gas in low-luminosity ellipticals is more vulnerable
to outflow than that in brighter ellipticals because the specific
gravitational binding energy decreases with galactic luminosity along
the fundamental plane.  Low-luminosity ellipticals are also thought to
be rotationally flattened which causes the X-ray luminosity of
interstellar gas to decrease sharply with increasing ellipticity.  In
our family of galaxies all with luminosity $L_B = 1.3 \times 10^{10}$
$L_{B,\odot}$ we find that $L_x$ is also very sensitive to the assumed
supernova rate.  In ellipticals having a ``low'' current supernova
rate, SNu$(t_n) = 0.01$, $L_x(t_n)$ varies along the sequence E0; E2; E6
in the ratios 1.0; 0.0091; 0.0027.  For a higher supernova rate,
SNu$(t_n) = 0.04$, the corresponding ratios are: 1.0; 0.025; 0.0033.
While the X-ray emission from hot interstellar gas in the E0 galaxies
would be easily observed with AXAF (by subtracting the stellar
component if that were possible), the prospects for observing the interstellar contribution
to $L_x$ for more realistic rotating galaxies of this $L_B$ are less
optimistic.  Furthermore, low luminosity E0 galaxies are rare or non
existent (Tremblay \& Merritt 1996).

The influence of past and current supernova rates on $L_x$ and other
important observational parameters is not entirely clear.  The higher
of the two current supernova rates we consider, SNu$(t_n) = 0.04$,
generates an iron abundance within the half-light radius that is 1.5
to 3 times the stellar iron abundance.  Of course any observation of
the X-ray iron features in rotating, low-luminosity ellipticals is
likely to be representative of stellar X-ray sources, not the gas. But
in more massive ellipticals the iron abundance in the interstellar gas
is often much {\it less} than that in the stars (Arimoto et al. 1997).
This could imply that the supernova iron -- and possibly also its
energy -- is not distributed throughout the interstellar medium as we
have assumed here; indeed the supernova energy is concentrated in hot
bubbles that may rise in the cooling flow atmosphere.  Because of this
as yet unsolved problem, we cannot be sure that we have treated the
supernova energy correctly in our models, particularly for the high
supernova rate solutions.  If our treatment of supernova energy is
correct, however, the disks of cool gas are severely truncated when
SNu$(t_n) \gta 0.04$ due to an enhanced outflow of high angular
momentum gas from the outer parts of the galaxy. The mass of the disk
remaining at time $t_n = 15$ Gyr $M_g(cold)$ is reduced by $\sim
1.7$ as SNu$(t_n)$ increases from 0.01 to 0.04. More luminous
ellipticals such as those we discussed in Paper 1 require much higher
SNu$(t_n)$ to alter the cold disks.

Of particular interest is the possibility of observing disks of cold
gas that are expected in rotating ellipticals either directly as
optical line emission or indirectly in the form of a younger stellar
population.  The column of unionized disk gas in our disks would
produce profound and observable absorption of soft X-rays, but this
can be avoided since gravitational clumping is expected in this
neutral or molecular gas.  We have estimated the density of ionizing
photons along the disk plane and computed the column density of warm
HII disk gas.  The presence of HII disk gas should be independent of
the likelihood of star formation in the colder neutral gas. For our
low value of SNu$(t_n)$, we find that large masses of HII gas can be
rotationally supported beyond $\sim 9$ kpc -- amounting to about one
tenth of all the gas ejected from galactic stars -- but the
density of this gas is too low to contribute to the observed optical
line flux.  In previous estimates of the total HII mass in ellipticals
a constant HII density is usually assumed; these masses may seriously
underestimate the true mass by several orders of magnitude.  In the
future we intend to investigate the survivability of gas in the outer
disk if the galaxy is moving through an external (inter-cluster)
medium. Most of the optical line emission in the disk comes from the
central parts of the disk which are less sensitive to rotation and the
supernova rate.  The total H$\beta$ luminosities expected from disk
HII are similar to those observed in ellipticals, implying that a
significant component of optical line emission from these galaxies may
originate in cooling flow disks.  If the disk HII emission is not
masked by other sources of warm gas, the optical line-emitting gas
should be (i) systematically flatter than the stellar image and (ii)
rotating at the circular velocity which exceeds the local mean line of
sight velocity of galactic stars.

We have found considerable support for the notion that stellar disks
in low luminosity ellipticals are a natural result of the evolution of
the interstellar gas.  The inevitability of cold disk formation in
galactic cooling flows is consistent with the finding that all low
luminosity ellipticals may contain stellar disks since many are hidden
due to low inclinations (Rix \& White 1990).  The stellar disk and
bulge are observed to share the same sense of rotation; this argues
against random merging events and supports star formation in cold
disks as we propose here. The total mass of gas that enters the cold
disk in 1 - 15 Gyrs, about 10 percent of the total stellar mass
$M_{*t}$, is very similar to the typical stellar mass observed in
elliptical disks.  Within the approximation of our galactic models
(e.g.  King and isothermal density profiles; constant Satoh $k^2$
factor, etc.) and that of current observations of stellar disks (most
of which have been studied only in very flat E5 or E6 ellipticals), we
believe that our non-exponential disks are similar to those observed.
Only three of the nine stellar disks studied by Scorza \& Bender
(1995) could be modeled with exponential disks and exponential fits
can always be made over a limited range in radius.  Finally, we have
demonstrated that the range of observed H$\beta$ photometric indices
observed in disky ellipticals is just spanned by our models if
luminous stars form in the cold gaseous disks.  We are implicitly
assuming here that the Type Ia supernova rate may vary among
ellipticals to generate a range of disk masses; at present there is no
compelling reason to believe otherwise. A similar variation of disk
masses could be obtained by beginning our calculation shortly before
or after 1 Gyr.  

If our ideas about star formation in disky
ellipticals are correct, these galaxies could become ideal
laboratories for studying successive generations of star formation. In
principle abundances in the parent population (bulge) stars, in the
interstellar gas that they have expelled, and in subsequent stellar
(disk) generations can all be directly observed.

While luminous star formation in disks provides a satisfactory
explanation for the ultimate fate of cooled gas in low luminosity
ellipticals, it is at present unclear why more luminous, slowly
rotating ellipticals do not also have visible stellar disks since cold
disks are expected there too (Paper 1).  Within the many uncertainties
regarding the star formation process, the stellar properties of all
elliptical disks can be understood if stars of lower mass are favored
in high pressure environments.  Maximum interstellar pressures in the
most massive ellipticals are about 1000 times that in the interstellar
medium of our Galaxy so even the most massive stars formed may not be
optically luminous; in low-luminosity ellipticals such as we consider
here the pressures are lower so stars at the upper IMF cutoff may be
optically luminous but insufficiently massive ($\lta 8$ $M_{\odot}$)
to produce SNII.  Alternatively, cold and dynamically fragile stellar
disks may have formed in luminous ellipticals and subsequently been
destroyed or altered perhaps in the same merging events responsible
for the overall boxy shapes of these galaxies.  Finally, hot ambient
(cluster) gas of low angular momentum may move into luminous
ellipticals and greatly reduce the size of cooling flow disks.  
We hope to address these problems in the near future.


\section{ACKNOWLEDGMENTS}

We are greatly indebted to Guy Worthey for preparing for us a version
of his dial-a-model program that can accommodate younger stars. 
The referee is thanked for helpful comments and corrections. Our
work on the evolution of hot gas in ellipticals is supported by grant
NAG 5-3060 of NASA to whom we are very grateful. In addition WGM is
partially 
supported by a UCSC Faculty Research Grant and FB is supported in 
part by Grant ASI-95-RS-152 from the Agenzia Spaziale Italiana.

\vskip.3in
\noindent
\centerline {\bf References}\\

\noindent
Arimoto, N., Matsushita, K., Ishimaru, Y., Ohashi, T.,
 \& Renzini, A. 1997, (preprint)\\
Binette, L., Magris, C. G., Stasinska, G., \& Bruzual, A. G. 1994,
 A\&A, 292, 13\\
Binney, J., Davies, R. L., \& Illingworth, G.D. 1990, ApJ, 361, 78\\
Brighenti, F. \& Mathews, W. G. 1996, ApJ, 470, 747 (Paper 1)\\
Brown, T. M., Ferguson, H. C., \& Davidsen, A. F. 1994, ApJ, 454,
 L15\\
Buson, L. M., Sadler, E. M., Zeilinger, W. W., Bertin, G. Bertola, F.,
 Danziger, J., Dejonghe, H., Saglia, R. P., \& de Zeeuw, P. T. 1993,
 AASup, 280, 409\\
Ciotti, L., D'Ercole, A., Pellegrini, S., \& Renzini, A., 1991, ApJ,
 376, 380\\
D'Ercole, A. \& Ciotti, L. 1997, private communication\\
Davies, R. L., Efstathiou, G., Fall, S. M., Illingworth, G.,
 \& Schechter, P. L. 1983, ApJ, 266, 41\\
Dehnen, W. 1993, MNRAS, 265, 250\\
de Jong, R. S. \& Davies, R. L. 1996,MNRAS (in print)\\
Dove, J. B. \& Shull, J. M., 1994, ApJ, 423, 196\\
Fabbiano, G., D.-W. Kim, \& Trinchieri, G. 1992, ApJS, 80, 531\\
 Faber, S. M., Tremaine, S., Ajhar, E. A., Byun Y-I.,
 Bressler, A., Gebhardt, K., Grillmair, C., Kormendy, J.,
 Lauer, T. R., \& Richstone, D. 1997, ApJ, (in press)\\
Faber, S. M., Trager, S. C., Gonzalez, J. J., \& Worthey, G., 1995,
 in {\it Stellar Populations}, eds. Gilmore, G., van der Kruit, P. C.,
 Proc. IAU Symp. 164, (Kluwer:Dordreht), 249\\
Gonzalez, J. J., 1993, Ph.D. thesis, Univ. of California at Santa
 Cruz\\
Hernquist, L. 1990, ApJ, 356, 359\\
Jaffe, W. 1983, MNRAS, 202, 995\\
Kim, D.-W., Fabbiano, G., \& Trinchieri, G. 1992, ApJ, 393, 134\\
Kormendy, J. \& Bender, R. 1996, ApJ, 464, L119\\
Macchetto, F., Pastoriza, M., Caon, N., Sparks, W. B., 
 Giavalisco, M., Bender, R. \& Capaccioli, M. 1996, A\&ASup, 
 120,463\\
Magris, G., \& Bruzual, G. 1993, ApJ, 417, 102\\
Mathews, W. G. 1989, AJ, 97, 42\\
Mathews, W. G., 1990, ApJ, 354, 468\\
Mathews, W. G., 1997, AJ, 113, 755\\
Nieto, J.-L., Bender, R., Arnaud, J., \& Surma, P. 1991, A\&A, 244,
 L25\\
Phillips, M. M., Jenkins, C. R., Dopita, M. A., Sadler, E. M., 
 \& Binette, L. 1986, AJ, 91, 1062\\
Rix, H.-W., \& White, S. 1990, ApJ, 362, 52\\
Satoh, C. 1980, Publ. Astr. Soc. Japan, 32, 41\\
Scorza, C. \& Bender, R. 1995, A\&A, 293, 20\\
Scorza, C. \& Bender, R. 1996, in {\it New Light on Galaxy Formation},
 eds. R. Bender \& R. L. Davies (Kluwer:Dordrecht), 55\\
Turatto, M., Cappellaro, E., \& Benetti, S. 1994,AJ, 108, 202\\
Tremblay, B. \& Merritt,D. 1996, AJ, 111, 2243\\
Tremaine, S., Richstone, D. O., Byun, Y.-I., Dressler, A,
 Faber, S. M., Grillmair, C., Kormendy, J., \& Lauer, T. R.
 1994, AJ, 107, 634\\
Trinchieri, G., \& di Serego Alighieri, SD. 1991, AJ, 101, 1647\\
Tsai, J. C. \& Mathews, W. G. 1995, ApJ, 448, 84\\
van der Marel, R. P. 1991, MNRAS, 253, 710\\
Worthey G., 1994, ApJS, 95, 107\\
Worthey, G., Trager, S. C., \& Faber, S. M., 1995, in {\it Fresh 
 Views of Elliptical Galaxies}, eds. Buzzoni, A., Renzini, A., 
 \& Serrano, A., ASP Conf. Series 86, 203\\
Zeilinger, W. W., Pizzella, A., Amico, P., Bertin, G., Bertola, F.,
 Buson, L. M., Danziger, I. J., Dejonghe, H., Sadler, E. M., 
 Saglia, R. P., \& de Zeeuw, P. T. 1996 AASup, 120, 257\\

\clearpage 

\begin{planotable}{l l}
\tablewidth{3.5in}
\tablenum{1}
\tablecaption{GALACTIC PARAMETERS FOR E0;0\tablenotemark{a}~ GALAXY}
\tablehead{
\colhead{Parameter} & 
\colhead{Value}\cr
}
\startdata

$R_{c*}$        &       63.65 pc\cr
$R_e$\tablenotemark{b}        &       1.725 kpc\cr
$R_t$           &       63.65 kpc\cr
$\rho_{*o}$     &       $2.373 \times 10^{-19}$~ gm cm$^{-3}$\cr
$M_{*t}$        &       $7.52 \times 10^{10}$ $M_{\odot}$\cr
$\beta = R_{ch}/R_{c*}$         &       31.623\cr
$\gamma = \rho_{ho}/\rho_{*o}$  &       $6.2448 \times 10^{-5}$\cr
$M_{ht}$        &       9 $M_{*t}$\cr
$L_B$           &       $1.317 \times 10^{10} L_{B\odot}$\cr
$M_B$		&	-19.89\cr
$M_V$           &       -20.83\cr
$M_{*t}/L_B$    &       5.71\cr
$\sigma_*$\tablenotemark{c}   &       291 km s$^{-1}$\cr

\tablenotetext{a}{A non-rotating ($k^2 = 0$) E0 galaxy.}
\tablenotetext{}{}
\tablenotetext{b}{Effective radius.}
\tablenotetext{}{}
\tablenotetext{c}{Characteristic velocity dispersion in stellar core,
$\sigma_* = ( 4 \pi G \rho_{*o} R_{c*}^2 /9 )^{1/2}$.}

\end{planotable}

\clearpage

\vskip.3in

\begin{planotable}{l l l l l l}
\tablewidth{6.0in}
\tablenum{2}
\tablecaption{GLOBAL COOLING FLOW EVOLUTION}
\tablehead{
\colhead{Galaxy}  &  
\colhead{Time}   & 
\colhead{$M_g (hot)$\tablenotemark{a}} & 
\colhead{$M_g (cold)$\tablenotemark{b}} &
\colhead{$L_x(\Delta E)$\tablenotemark{c}} & 
\colhead{$R_d$\tablenotemark{d}} \nl
\colhead{     }   & 
\colhead{(Gyr)}  & 
\colhead{($10^{9} M_{\odot}$)}  & 
\colhead{($10^{9}M_{\odot}$)} & 
\colhead{($10^{39}$ ergs/s)} & 
\colhead{(kpc)}
}

\startdata

E0L;0   &  4  &  0.74   & 3.98   & 121.8 & ... \cr
        & 10  &  0.59   & 6.41   & 34.6 & ... \cr
        & 15  &  0.49   & 7.38   & 19.8 & ... \cr
        &     &         &        &      &      \cr
E0H;0   &  4  &  0.76   & 2.77   & 72.5 &  ...   \cr
        & 10  &  0.49   & 4.17   & 14.1 &  ...   \cr
        & 15  &  0.32   & 4.57   & 5.10 &  ...   \cr
        &     &         &        &      &      \cr
E2L;1   &  4  & 0.65   & 4.03   & 3.30   & $\sim$7.5 \cr
        &  10 & 0.40   & 6.47   & 0.50   & $\sim$13.5 \cr
        & 15  & 0.14   & 7.55   & 0.18   & $\sim$31.5 \cr
        &     &         &        &      &      \cr
E2H;1   &  4  &  0.74  & 2.78   & 7.48  & $\sim$0.75  \cr
        &  10 &  0.44  & 4.14   & 2.37  & $\sim$0.75  \cr
        & 15  &  0.29  & 4.40   & 0.13  & $\sim$0.65  \cr
        &     &         &        &      &      \cr
E6L;1   &  4  & 0.53  & 4.11  & 1.10  & $\sim$27  \cr
        &  10 & 0.16  & 6.55  & 0.16  & $\sim$42.5  \cr
        & 15  & 0.08  & 7.63  & 0.053 & $\sim$52  \cr
        &     &         &      &      &       \cr
E6H;1   &  4  &  0.67   & 2.72  &  2.34  &  $\sim$2.3   \cr
        &  10 &  0.37   & 4.13  &  0.20  & $\sim$4.5    \cr
        & 15  &  0.08   & 4.40  &  0.017 & $\sim$3.8    \cr

\tablenotetext{a}{Total gas mass in the galaxy
having temperature $>$ $10^4$K.}

\tablenotetext{b}{Total gas mass of cooled gas within galaxy with
temperature $T = 10^4$K.}

\tablenotetext{c}{Xray luminosity in the ROSAT band.}

\tablenotetext{d}{Disk radius.}

\end{planotable}

\clearpage

\begin{planotable}{l l l}
\tablewidth{3.8in}
\tablenum{3}
\tablecaption{STELLAR BURSTS IN THE DISK AND BULGE}
\tablehead{
\colhead{     }  &  
\colhead{E2L;1}   &
\colhead{E2H;1} \nl
\colhead{$~$}  &
\colhead{$~$}  &
\colhead{$~$}  \nl
\colhead{$t_i$\tablenotemark{a}} &
\colhead{$\Delta M(t_i)$\tablenotemark{b}} & 
\colhead{$\Delta M(t_i)$\tablenotemark{b}} \nl
\colhead{(Gyr)}  & 
\colhead{($10^{8} M_{\odot}$)}  & 
\colhead{($10^{8}M_{\odot}$)}  \nl
}

\startdata

15.0\tablenotemark{c}  &  131.1\tablenotemark{c}  
&  131.1\tablenotemark{c}   \cr
12.5                   &  28.56  &   27.80  \cr
9.5                    &   8.24  &    9.00  \cr
6.5                    &   4.44  &    4.61  \cr
3.5                    &   2.78  &    2.02  \cr
1.7                    &   0.43  &    0.18  \cr
1.1                    &   0.43  &    0.18  \cr
0.4                    &   0.58  &    0.24  \cr

\tablenotetext{a}{Age of burst.}

\tablenotetext{b}{Mass of burst within $R_e/2$.}

\tablenotetext{c}{Age and mass of galactic bulge within $R_e/2$.}

\end{planotable}


\begin{planotable}{l l l}
\tablewidth{3.0in}
\tablenum{4}
\tablecaption{H$\beta$ INDICES\tablenotemark{a} ~ WITHIN $R_e/2$ }
\tablehead{
\colhead{     }  &  
\colhead{E2L;1}   &
\colhead{E2H;1} \nl 
}

\startdata

Galaxy alone    &  1.41  &  1.41   \cr
Disk            &  2.87  &  2.29   \cr
Galaxy + disk   &  2.03  &  1.73   \cr

\tablenotetext{a}{Using Padova evolution tracks.}

\end{planotable}

\clearpage 

\noindent
\centerline{\bf Figure Captions}\\

\noindent
{\bf Figure 1:} Contours of the stellar
temperature $\log T_*(R,z)$ (light lines) and azimuthal stellar
velocity $v_{* \phi}(R,z)$ (heavy lines) for the E2;1 galaxy.
Values of $\log T_*(R,z)$ and $v_{* \phi}(R,z)$ for
two contours are labeled (units: K and k s$^{-1}$);
the contours are equally spaced in $\log T_*$ and $v_{* \phi}$.

\noindent
{\bf Figure 2:} Variation of the circular velocity $v_{circ}(R)$
(solid lines)
and stellar azimuthal velocity $v_{* \phi}(R,0)$ on the equatorial
plane in the E2;1 (bold lines) and E6;1 (light lines) galaxies.
All velocities in km s$^{-1}$.

\noindent
{\bf Figure 3:} Properties of the E0;0 galaxy (dashed lines) 
and its interstellar
gas after 15 Gyr for two supernova rates:
E0L;0 (heavy solid lines) and E0H;0 (light solid lines). 
a) Variation of stellar density
$\rho_*(R)$ and gas density $\rho(R)$ (gm cm$^{-3}$);
b) Variation of stellar temperature $T_*(R)$
and gas temperature $T(R)$ 
c) Distribution of 0.5 - 4.5 keV X-ray surface brightness
$\Sigma_x(R)$ (solid lines)
and stellar surface
brightness $\Sigma_*(R)$ (dashed line) with projected radius;
the vertical normalization is arbitrary.

\noindent
{\bf Figure 4:} Plots of the surface density in the
disk of cold gas $\Sigma_d(R,t)$ (gm cm$^{-2}$)
as a function
of radius $R$ at three times: $t = 4$ Gyr (dot-dashed line),
10 Gyr (dashed line) and 15 Gyr (light solid line)
for the a) E2L;1 galaxy, 
b) E2H;1 galaxy, and c) E6L;1 galaxy.

\noindent
{\bf Figure 5:} Equally spaced
contours of the logarithm of the ultimate disk radius
in parsecs, $\log R_d(R,z)$, for gas ejected from stars throughout
the E2L;1 galaxy.
Two contours are labeled with values of $\log R_d(R,z)$
in pc.

\noindent
{\bf Figure 6:} Four panels describe the interstellar flow 
in galaxy E2L;1 at time $t_n = 15$ Gyr; all contours are
equally spaced.
The physical limit of the galaxy is shown with a dashed line.
{\it upper left}: Contours of $\log \rho(R,z)$ with two 
contours labeled. Velocity arrow at upper left is 50 km s$^{-1}$;
{\it upper right}: contours of $\log \rho(R,z)$ in the 
central galaxy with two contours labeled.
Velocity arrow at upper left is 50 km s$^{-1}$;
{\it lower left}: Contours of $\log T(R,z)$ with two labeled;
{\it lower right}: Contours of azimuthal velocity 
$v_{\phi} (R,z)$ with two contours labeled in km s$^{-1}$.

\noindent
{\bf Figure 7:} Flow and thermal conditions in the E2H;1 galaxy
at $t_n = 15$ Gyr; all contours are
equally spaced.
The physical limit of the galaxy is shown with a dashed line.
{\it upper panel}: Contours of $\log \rho(R,z)$ with two labeled.
Velocity arrow at upper left is 50 km s$^{-1}$;
{\it lower panel}: Contours of $\log T(R,z)$ with two labeled.

\noindent
{\bf Figure 8:} Gas flow in the E6L;1 galaxy
at $t_n = 15$ Gyr; all contours are
equally spaced.
The physical limit of the galaxy is shown with a dashed line.
Two contours of $\log \rho(R,z)$ are labeled.
Velocity arrow at upper left is 50 km s$^{-1}$.

\noindent
{\bf Figure 9:} X-Ray surface brightness $\Sigma_x(R,z)$
(0.5 - 4.5 keV) distribution
in three galaxies after $t = 15$ Gyr viewed
perpendicular to the axis of rotation.
Two contours are labeled with values of $\log \Sigma_x$
in erg s$^{-1}$ cm$^{-2}$;
all contours are equally spaced.
{\it Top}: $\Sigma_x(R,z)$ for galaxy E2L;1;
{\it middle}: $\Sigma_x(R,z)$ for galaxy E2H;1;
{\it bottom}: $\Sigma_x(R,z)$ for galaxy E6L;1.

\noindent
{\bf Figure 10:} Column densities along the disk plane of the 
E2L;1 galaxy at time $t_n = 15$ Gyr.
{\it Dashed line:} the total vertical column of cooled gas $N_{tot}$;
{\it solid line:} the maximum column $N_i$ that can be photoionized HII gas.
The photoionized column at any radius is 
$N_{HII} = {\rm min} (N_i,N_{tot})$.

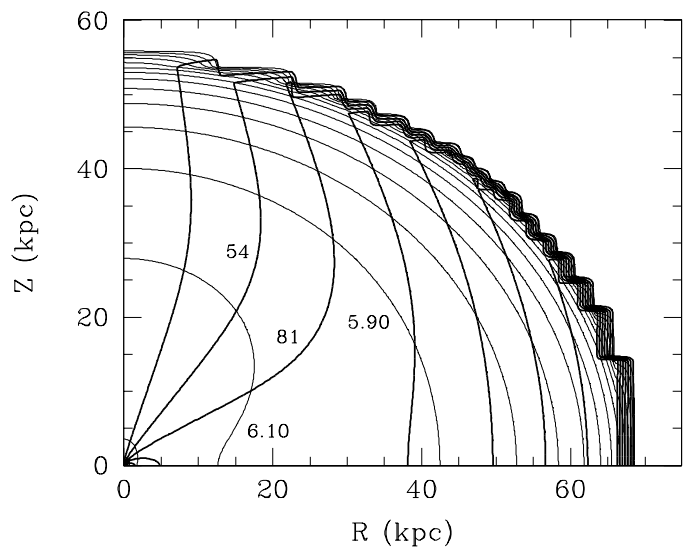
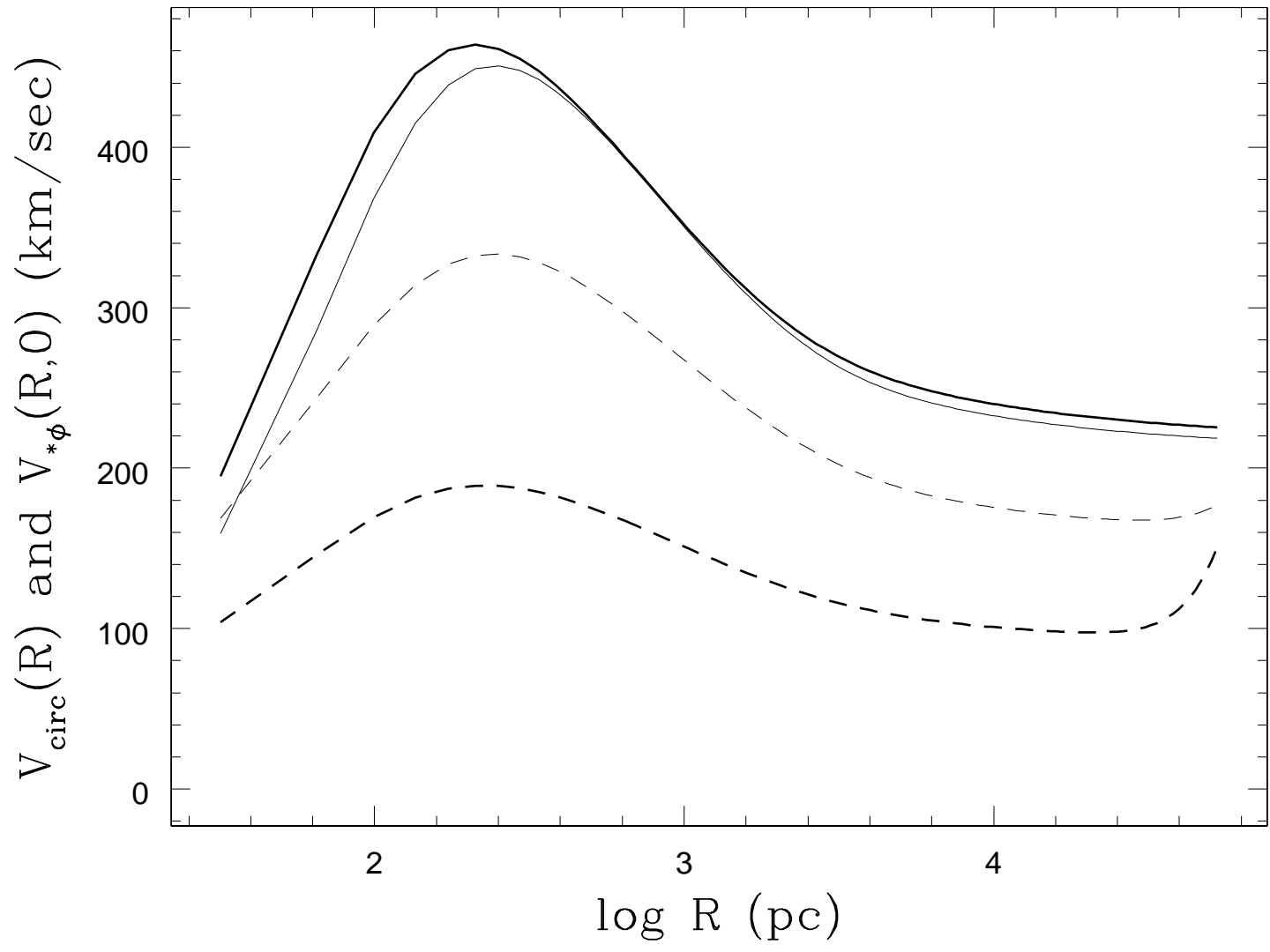
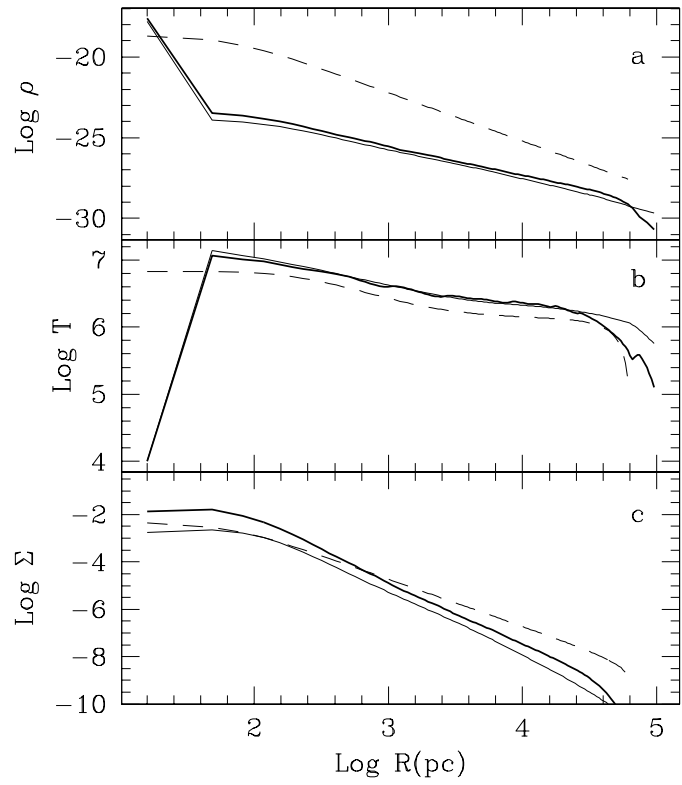
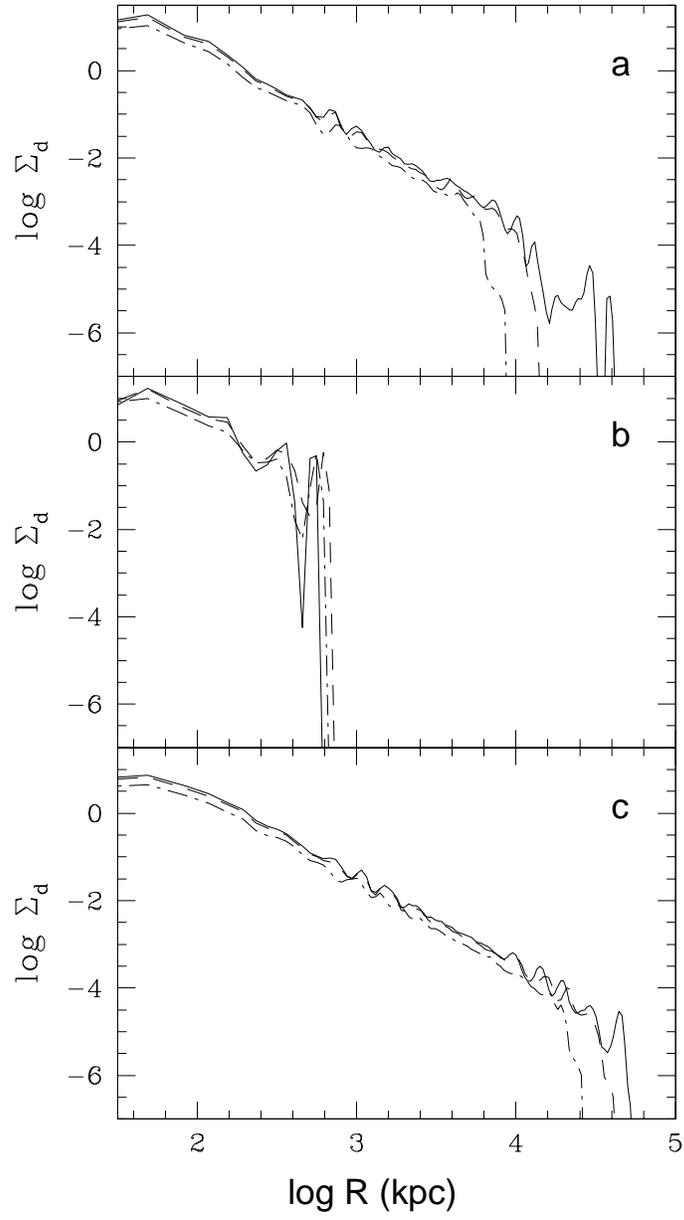
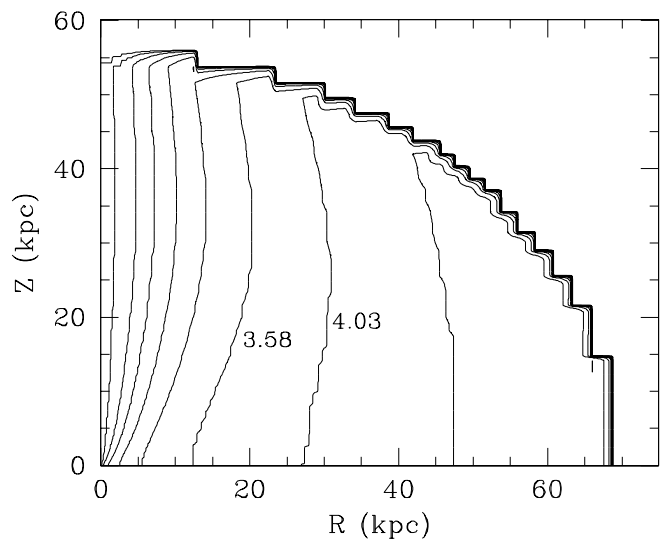
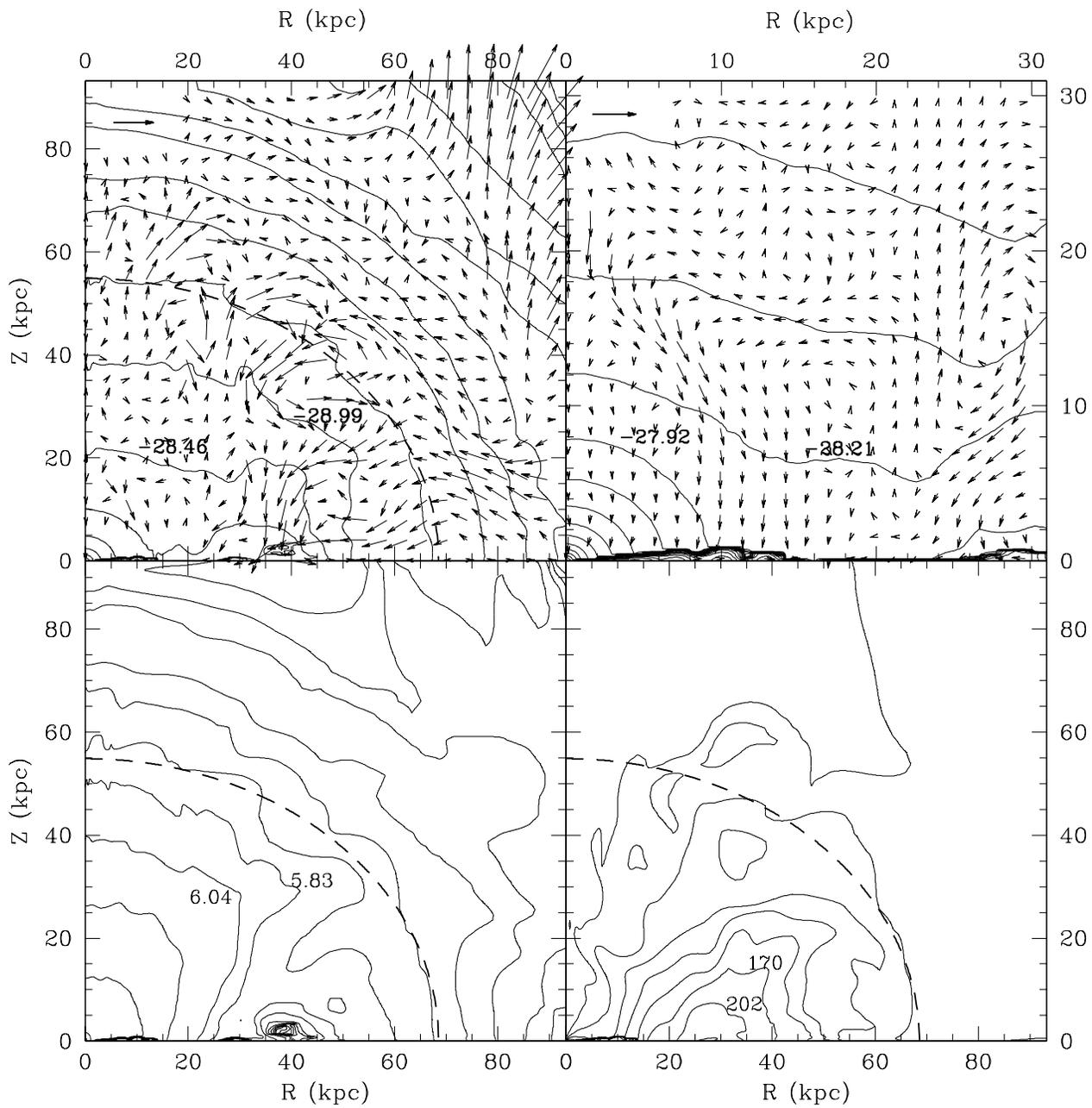
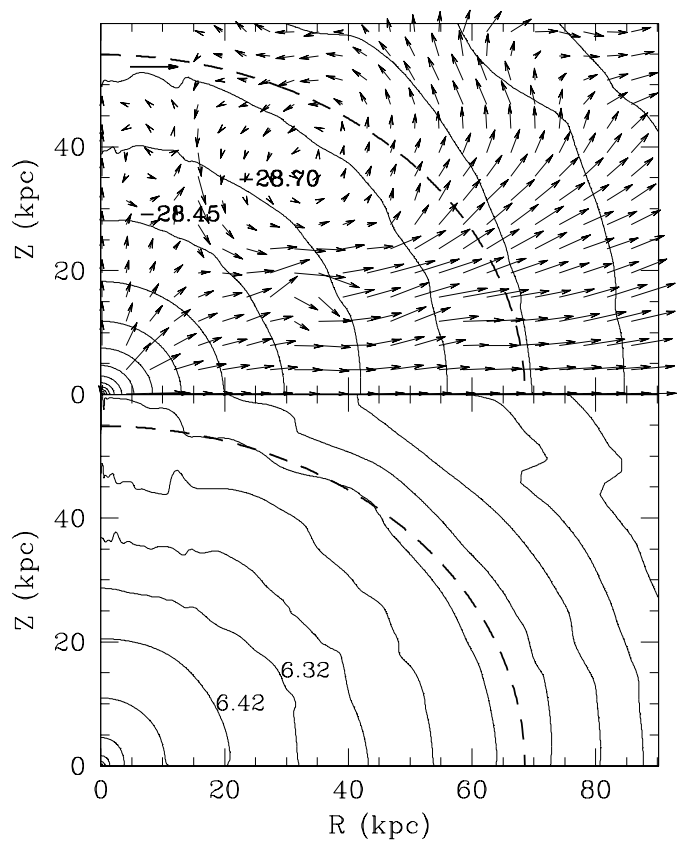
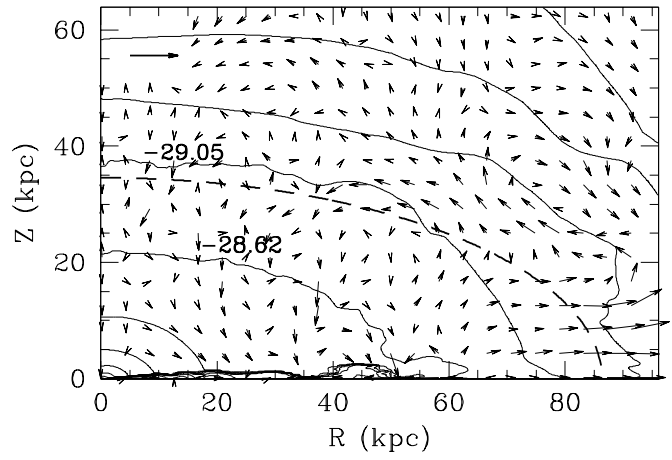
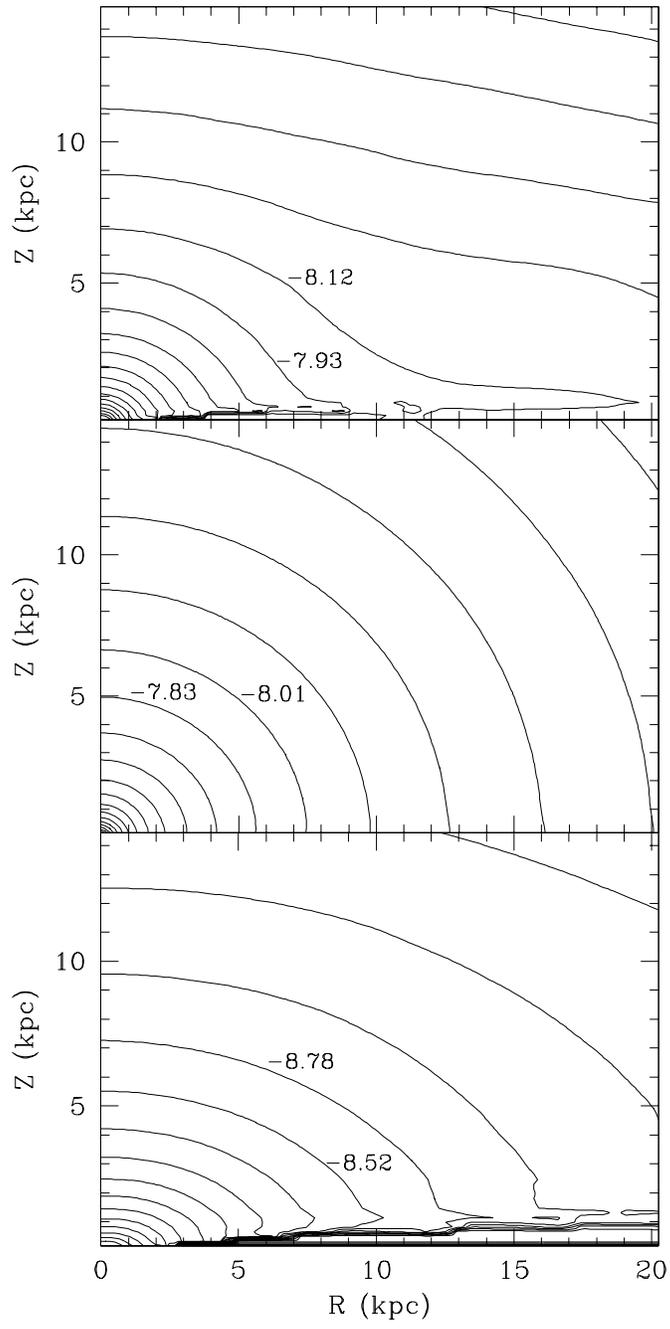
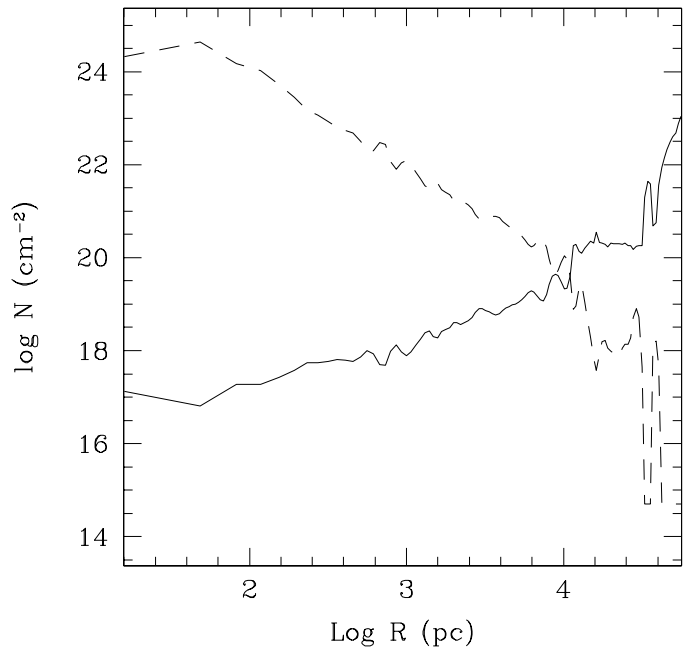
\end{document}